\documentclass[pre,showpacs,twocolumn,preprintnumbers,amsmath,amssymb,showkeys]{revtex4}

\usepackage{hyperref}   
\usepackage{graphicx}   
\usepackage{amsmath,amssymb}

\renewcommand{\vec}{\mathbf}
\newcommand{\vvec}{\vec{v}}

\newcommand{\vorvec}{\vec{\omega}}
\newcommand{\sfunc}{\psi}

\newcommand{\R}{{\rm Re}}
\newcommand{\Dean}{{\rm De}}

\newcommand{\Jacobian}[2]{J\left({#1},{#2}\right)}

\renewcommand{\phi}{\varphi}

\newcommand{\pd}[2]{\frac{\partial{#1}\mathstrut}{\partial{#2}}}

\newcommand{\pdd}[2]{\frac{\partial^2\!{#1}}{\partial {#2}^2}}

\DeclareMathOperator{\rot}{\nabla\times}

\DeclareMathOperator{\Lapl}{\triangle}
\begin{document}

\title{Full perturbation solution for the flow in a rotating torus}%
\author{A. Chupin}%
\email{chupin@icmm.ru (corresponding author)}%
\author{R. Stepanov}
\affiliation{Institute of Continuous Media Mechanics, Korolyov 1, Perm, 614013, Russia}

\date{\today}%
\begin{abstract} We present a perturbation solution for a pressure-driven
fluid flow in a rotating toroidal channel. The analysis shows the
difference between the solutions of full and simplified equations
studied earlier. The result is found to be reliable for {\it low}
Reynolds number ($\R$) as was the case for a previously studied
solution for high $\R$. The convergence conditions are defined for
the whole range of governing parameters. The viscous flow exhibits
some interesting features in flow pattern and hydrodynamic
characteristics.
\end{abstract}
\maketitle
\section{Introduction}

Fluid flow in a general curved pipe is known not to be
one-dimensional since the pioneering work of Dean \cite{Dean1927}. A
secondary flow develops when a fluid is driven by pressure gradient
in a toroidal channel. In practice, this occurs in many
applications: channels in industry, blood vessels in physiology,
transport pipe systems, coolant pipe systems and others. In a
considerable number of studies there has been a strong motivation
for finding a solution for flows in a curved channel.

A stationary solution of the Navier-Stokes equation in a curved pipe
was first analyzed by the perturbation method \cite{Dean1927}. An
analytical solution was obtained for a {\it simplified} equation
valid at a low value of the Dean number $\Dean=\R \, \kappa^{1/2}$.
This solution was compared with the results of direct numerical
simulation for the wide range of governing parameters
\cite{zhang:056303}. Later many authors considered this problem
using different approaches. Pressure drop, heat and mass transfers
were studied in detail (for review see  e.g. \cite{zhang:056303}).
Non-Newtonian fluid flow in a curved pipe shows a specific behavior
for different rheologic cases (e.g. \cite{2006PhFl...18h3103C} and
references within). Magnetic field induction was experimentally
studied in conductive fluid flow in toroidal channel
\cite{2006PhRvE..73d6310S}.

In the present paper we derive a perturbation analytical solution of
the {\it full} equations taking into account all curvature effects.
It is shown that the well known solution \cite{Dean1927} corresponds
to the case $\R\gg1$. In the general case the solution can not be
parameterized by the Dean number only. The suggested solution
properly describes the case of low Reynolds number ($\R$) and
asymptotically approaches the known solution at high $\R$. For high
viscosity of the fluid the inertial terms in the Navier-Stokes
equation are relatively unimportant and the flow pattern is
determined by a balance of  viscous forces and the pressure
gradients in the fluid. Such flows, called creeping flows, are of
nearly the same importance in practice as the inertial flows (with
$\R\gg 1$).

\section{\label{sec:1}Mathematical model}

\begin{figure}
\includegraphics[width=5cm]{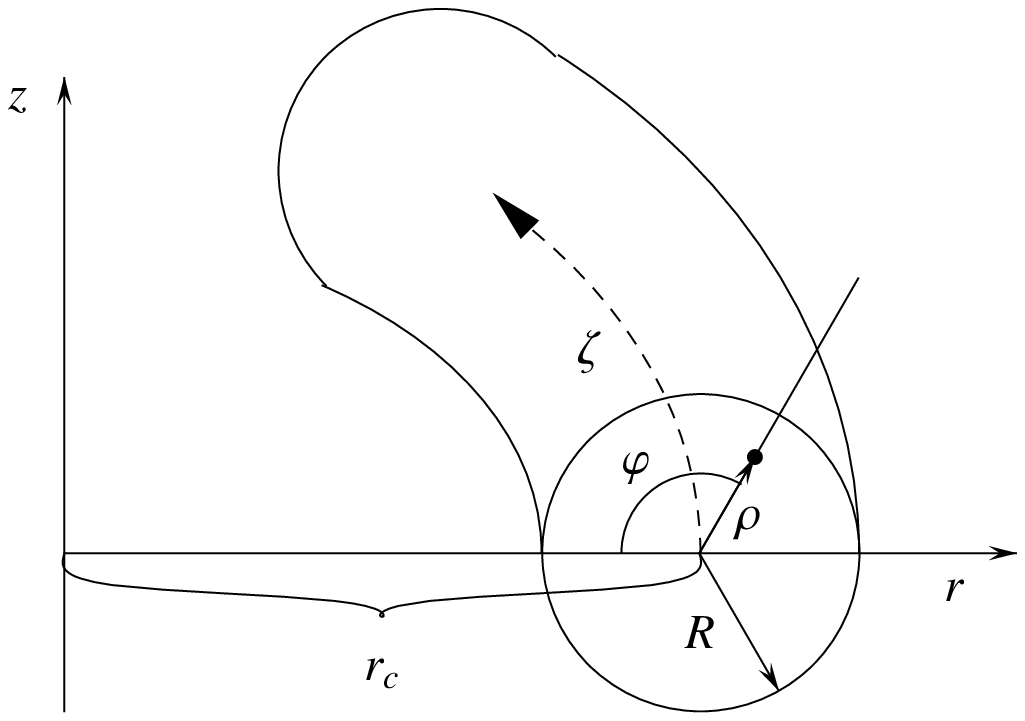}
\caption{The coordinate system in the toroidal channel.}
\label{pic:sys}
\end{figure}

A curved pipe is considered as a toroidal channel with the outer
radius $R_c$ and radius of the inner circular section $R$. We use a
coordinate system $\{\rho, \phi, \zeta\}$, where $\rho$ and $\phi$
are polar coordinates in the cross-section and $\zeta$ is the linear
coordinate along  the channel (see Fig.~\ref{pic:sys}).
The torus revolves around its main axis with a constant angular
velocity $\Omega$ ($\Omega>0$ means co-rotation). The flow of an
incompressible fluid in the rotating channel is governed by the
equations
\begin{align}\label{NS}
\begin{split}
\pd{\vvec}{t}+\vvec\cdot\nabla\vvec&=-2\vec{\Omega}\times\vvec-\frac{\nabla p}{\rho_0} + \nu \triangle \vvec,\\
\nabla\cdot\vvec&=0,
\end{split}
\end{align}
where $\vvec$ is the velocity, $\rho_0$ is the fluid density and
$\nu$ is the kinematic viscosity.

After introducing the vorticity $\vorvec=\nabla\times\vvec$ pressure
can be eliminated. The following {\it Helmholtz equation} is written
as
\begin{equation}\label{Helm}
\pd{\vorvec}{t}+\left(\vvec\cdot\nabla\right)\vorvec-\left(\vorvec\cdot\nabla\right)\vvec=-2\nabla\times(\vec{\Omega}\times\vvec)
+ \nu \Lapl\vorvec.
\end{equation}
Assuming that the flow is stationary and $\zeta$-independent
Eqs.~\eqref{NS}-\eqref{Helm} can be reduced to the equations for two
scalar functions: stream velocity $v_\zeta$ and stream function
$\sfunc$ which is defined in such way that
\begin{equation}\label{sfunc-def}
\left\{v_\rho,v_\phi,0\right\}=\rot\left(\sfunc\hat{\vec{\zeta}}\right)
\end{equation}
The no-slip boundary conditions
\begin{equation}\label{BC}
v_\zeta|_{\rho=1}=\sfunc|_{\rho=1}=\left.\pd{\sfunc}{\rho}\right|_{\rho=1}=0
\end{equation}
together with the regularity condition at the centerline must be
satisfied. Following arguments in \cite{1968RSPSA.307...37M} we
assume the pressure to be a linear function of $\zeta$. Therefore
the pressure gradient is constant vector along channel.

Choosing $V$ as the characteristic velocity, the inner radius $R$ as
the characteristic  length  and $Q$ as the characteristic pressure
gradient we introduce dimensionless variables $\rho^*=\rho/R, \,
\zeta^*=\zeta/R$, $ v_\zeta^*=v_\zeta/V,\, \sfunc^*=\sfunc/(V R)$,
$p^*=p R/K$. The following governing dimensionless parameters will
be used: curvature ratio $\kappa=R/R_c$, Reynolds number $\R=(V
R)/\nu$, dimensionless angular velocity $F=(\Omega r_c)/V$ and
pressure parameter $G=(Q R^2)/(\rho_0 \nu V)$ which is a ratio of
the viscous and pressure forces. $r=(\kappa^{-1}- \rho \cos\phi)$ is
the dimensionless distance from the torus axis.

Thus, the following set of equations describes the problem (the
superscript '*' is omitted everywhere below)
\begin{eqnarray}\label{eqvzeta}
 \left(D_\phi\sfunc D_r v_\zeta - D_r\sfunc D_\phi v_\zeta\right) = \nonumber \\
 2\kappa\,F\left(\frac{\cos\phi}{\rho}\pd{\sfunc}{\phi} +
\sin\phi\pd{\sfunc}{\rho}\right) + \frac{G}{r \kappa {\R} } +
\frac{1}{\R}\mathcal{L} v_\zeta, \\ \label{eqsfunc} D_r v_\zeta
\hat{D}_\phi v_\zeta
    - \hat{D}_rv_\zeta  D_\phi v_\zeta
+ D_r\sfunc \hat{D}_\phi\mathcal{L}\sfunc
  - \hat{D}_r\mathcal{L}\sfunc D_\phi\sfunc = \nonumber\\
 2\kappa F\left(\frac{\cos\phi}{\rho}\pd{v_\zeta}{\phi} +
\sin\phi\pd{v_\zeta}{\rho}\right)
   -\frac{1}{\R}\mathcal{L}^2\sfunc,
\end{eqnarray}
while the derivation of the curl in these coordinates leads to the
expression for stream function:
\begin{equation}\label{DrDf}
\begin{split}
v_\rho&=D_\phi\sfunc=\frac{\sfunc\sin\phi}{\kappa^{-1}-\rho\cos\phi}+\frac{1}{\rho}\pd{\sfunc}{\phi},\\
v_\phi&=-D_r\sfunc=\frac{\sfunc\cos\phi}{\kappa^{-1}-\rho\cos\phi} - \pd{\sfunc}{\rho}.
\end{split}
\end{equation}

Other differential operators are
\begin{eqnarray}
\hat{D}_r&=&\pd{}{\rho} + \frac{\cos\phi}{r}, \qquad \label{op:1}
\hat{D}_\phi=\frac{1}{\rho}\pd{}{\phi} - \frac{\sin\phi}{r},\\
\Lapl_{cyl}&=&\pdd{}{\rho} + \frac{1}{\rho}\pd{}{\rho}+
\frac{1}{\rho^2}\pdd{}{\phi} \label{op:cyllapl}\\
\mathcal{L}&=&\Lapl_{cyl}- \frac{\cos\phi}{r}\pd{}{\rho} -
\frac{1}{r^2} + \frac{\sin\phi}{\rho r}\pd{}{\phi}.
\label{op:torlapl}
\end{eqnarray}

The effect of curvature can be obtained after significant
simplification of the governing equations (\ref{eqvzeta}) and
(\ref{eqsfunc}). The idea of Dean \cite{Dean1927} was to renormalize
the stream function by a factor of $\R$ and combine factors $\R \,
\kappa^{1/2}$ into the so called Dean number. The approximation
$\kappa\ll 1$ is applied after that. Terms due to curvature remain
only in the first two advective terms and the Coriolis force term in
Eq.~(\ref{eqsfunc}). Limit $\kappa\ll 1$ and fixed value $\Dean$
corresponds to limit $\R\gg 1$. A perturbation solution of these
{\it simplified} equations was reconsidered and compared with the
results of direct numerical simulations in \cite{zhang:056303}. We
solve the equations (\ref{eqvzeta}) and (\ref{eqsfunc}) in general.
This means that the viscous term corresponding to the toroidal
Laplacian operator $\mathcal{L}$ (\ref{op:torlapl}) is not replaced
by a cylindrical one (\ref{op:cyllapl}). The term corresponding to
the Coriolis force in Eq.~(\ref{eqvzeta}) remains. All curvature
contributions are kept in the advective terms and pressure gradient.

\section{\label{sec:2}Perturbation solution}
Solution of Eqs.~(\ref{eqvzeta}) and (\ref{eqsfunc}) can be expanded
in power series in a small parameter $\kappa$
\begin{eqnarray}
\label{series}
v_\zeta &= v^{(0)}(\rho,\phi) + \kappa v^{(1)}(\rho,\phi) + \kappa^2 v^{(2)}(\rho,\phi) + \ldots,\\
\sfunc &= \sfunc^{(0)}(\rho,\phi) + \kappa \sfunc^{(1)}(\rho,\phi) +
\kappa^2 \sfunc^{(2)}(\rho,\phi) + \ldots
\end{eqnarray}
Expanding all operators (\ref{DrDf})-(\ref{op:torlapl}) in the
same series we can write the zero-order approximation of
Eqs.~(\ref{eqvzeta}) and (\ref{eqsfunc}) as
\begin{equation}\label{eq0}
\begin{split}
\Jacobian{v^{(0)}}{\sfunc^{(0)}}
    &= \frac{G}{\R}+\frac{1}{\R}\Lapl_{cyl}v^{(0)},\\
\!\Jacobian{\Lapl_{cyl}\sfunc^{(0)}}{\sfunc^{(0)}}&=
\frac{1}{\R}\Lapl_{cyl}^2\sfunc^{(0)},
\end{split}
\end{equation}
where $\Jacobian{A}{B}$ is the Jacobian of the functions $A$ and $B$ with respect to $\rho$ and $\phi$:
\begin{equation}\label{jacobian}
\Jacobian{A}{B}=\frac{1}{\rho}\left(\pd{A}{\rho}\pd{B}{\phi}-\pd{B}{\rho}\pd{A}{\phi}\right).
\end{equation}
This gives a solution corresponding to cylindrical geometry, namely to the
Poiseuille flow
\begin{equation}
\label{sol0} v^{(0)}=\frac{G}{4} \left(1-\rho ^2\right), \;
\sfunc^{(0)}=0.
\end{equation}
The  first-order approximation leads to the system
\begin{multline}\label{eq1v}
\Jacobian{v^{(1)}}{\sfunc^{(0)}} + \Jacobian{v^{(0)}}{\sfunc^{(1)}}+ \sfunc^{(0)}v^{(0)}\pd{}{z}\ln\frac{v^{(0)}}{\sfunc^{(0)}} + \\
+ \frac{1}{\R}\pd{v^{(0)}}{r} = \frac{G}{\R}\rho\cos\phi + \frac{1}{\R}\Lapl_{cyl}v^{(1)}
+ 2F\pd{\sfunc^{(0)}}{z},
\end{multline}
\begin{multline}\label{eq1sf}
\frac{\Lapl_{cyl}^2\sfunc^{(1)}}{\R}=
\Jacobian{\Lapl_{cyl}\sfunc^{(0)}}{\sfunc^{(1)}} + \Jacobian{\Lapl_{cyl}\sfunc^{(1)}}{\sfunc^{(0)}} +\\
  +2\left(F+v^{(0)}\right) \pd{v^{(0)}}{z} - \frac{2}{\R}\pd{}{r} \Lapl_{cyl}\sfunc^{(0)} +\\
  +\Jacobian{\pd{\sfunc^{(0)}}{r}}{\sfunc^{(0)}} + \pd{\left(\sfunc^{(0)}\cdot\Lapl_{cyl}\sfunc^{(0)}\right)}{z}
\end{multline}
and its solution gives the first-order correction
\begin{multline}\label{sol1v}
v^{(1)}=\frac{3G\cos\phi}{16}\rho\left(1-\rho ^2\right)+\\
+ \frac{G^3\R^2\cos\phi}{737280} \rho \left(1-\rho ^2\right)\left(\rho^6-9\rho^4+21\rho ^2-19\right)-\\
- \frac{G^2 F\R^2\cos\phi}{18432} \rho  \left(1-\rho ^2\right) \left(\rho^4-3\rho^2+3\right),
\end{multline}
\begin{multline}\label{sol1sf}
\sfunc^{(1)}=\frac{G^2\R \sin\phi}{4608}\rho  \left(\rho ^2-4\right) \left(\rho ^2-1\right)^2-\\
-\frac{G F\R\sin\phi}{192} \rho  \left(\rho ^2-1\right)^2.
\end{multline}
This solution is different from the solution obtained by
\cite{zhang:056303} (we denote the latter by superscript 'c'). The
first-order solution has an additional term
\begin{align}\label{new_v_term}
v_\zeta-v_\zeta^{(c)}&=\kappa\frac{3G\cos\phi}{16}\rho\left(1-\rho
^2\right)+O\left(\kappa^2\right).
\end{align}
$\sfunc^{(1)}$ has no difference from that in \cite{zhang:056303}.
The corrections of the second-order approximation $v^{(2)}$ and
$\sfunc^{(2)}$ can be derived in a similar way. For the stream
function we find
\begin{align}\label{new_sfunc_term}
\sfunc-\sfunc^{(c)}&=-\kappa^2\frac{\rho^2 \left(1-\rho ^2\right)^2 \sin 2\phi}{92160} \left(150 G \text{Re}F + \right.\nonumber \\
&\left.+ G^2 \text{Re} \left(56 - 17 \rho ^2\right)
\right)+O\left(\kappa^3\right).
\end{align}
We note that $\kappa$ and $\R$ cannot be combined (for instance, in
the Dean number) for parametrization of the present perturbation
solution, both should be treated as independent parameters.

\section{\label{sec:3}Convergence}
It is clear that the perturbation solution is valid only at specific
values of the governing parameters $\R$, $F$, $G$ and $\kappa$ and
the series must converge. The conditions on the governing parameters
have been defined in order to satisfy the uniform convergence of the
maximum value for each order of approximation.

All terms in the expressions for $v^{(i)}$ and $\sfunc^{(i)}$ have
the form $(G\R)^{j}(F\R)^{k}/\R$. So, for convergence analysis we
use for convenience the following values: $G'=G\R$ and $F'=F\R$
which are the only factors that influence the convergence. The
maximal powers of $G'$ and $F'$ included in $v^{(i)}$ grow as $2i+1$
and $i$ accordingly. However, these powers can be reached only in
different terms such as $G'^{2i+1}$ and $G'^{i+1}F'^{i}/\R$. So the
radius of convergence should be $O(G'^{-2} + F'^{-1})$. We estimate
the radius of convergence $\kappa_c$ with the following technique. A
supposition of uniform convergence neccessarily implies a monotonic
decrease of $v^{(i)}$ majorants in \eqref{series}. In order to know
this we evaluate terms up to the 5th order, find the global maxima
in the circle $\rho\le 1$, and calculate largest $\kappa$, for which
the maxima do not increase.

\begin{figure}
\includegraphics[width=0.32\textwidth]{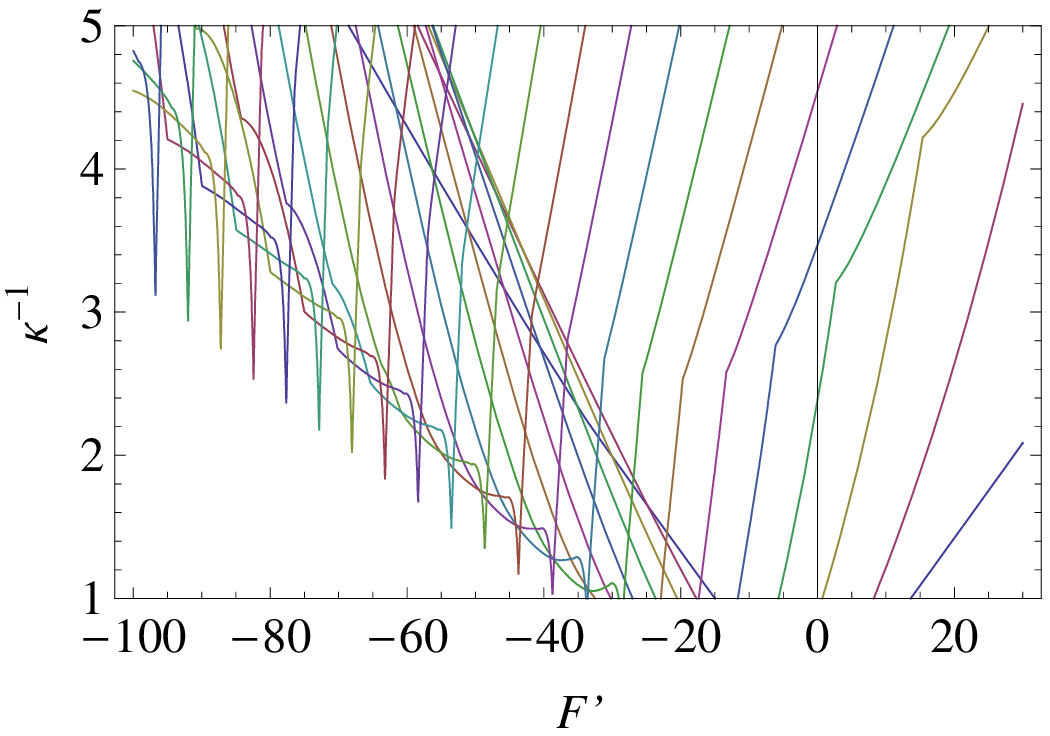}
\includegraphics[width=0.32\textwidth]{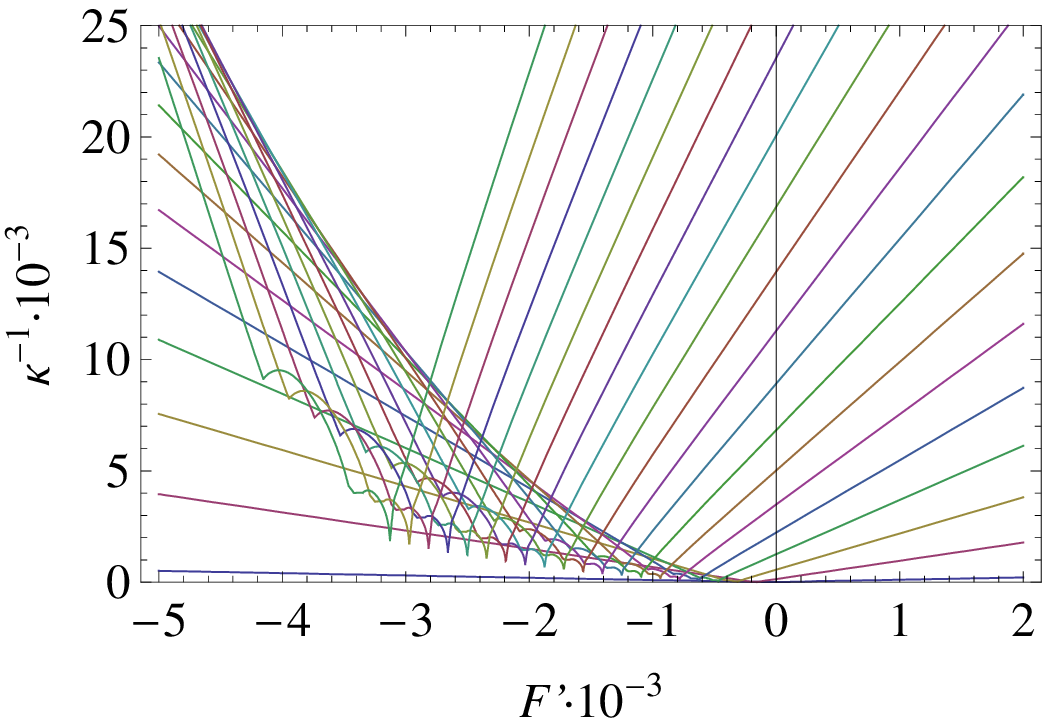}
\caption{The dependence of convergence radius on $F'$ for $G' =
5..705$ (top) and for $G' = 5..21000$ (bottom).} \label{fig_conv}
\end{figure}

Thus we obtain $\kappa_c$ as a function of $G'$ and $F'$ (see
Fig.~\ref{fig_conv}), rather complex but holding some nice
asymptotics. For large $F'$ we have obtained a dependance of the
following kind: $a+b\left|F'-c\right|$. As pattern fitting has
shown, the coefficients $a$, $b$ and $c$ grow linearly in $G'$. The
slope of the infinite branches grows as $b\sim G'$.

\section{\label{sec:4}Results}

Detailed analysis of the solution was done in the earlier studies.
Where the attention was centered on the case of large $\R$. Our new
finding deals with the case of small $\R$ when the viscous term is
comparable or larger than the convective one. This difference arises
when new terms in the solution
(\ref{new_v_term}-\ref{new_sfunc_term}) dominate, i.e. when $\R$ is
small.

\begin{figure}
\begin{picture}(200,200)(0,0)
\put(0,100){\includegraphics[width=0.2\textwidth]{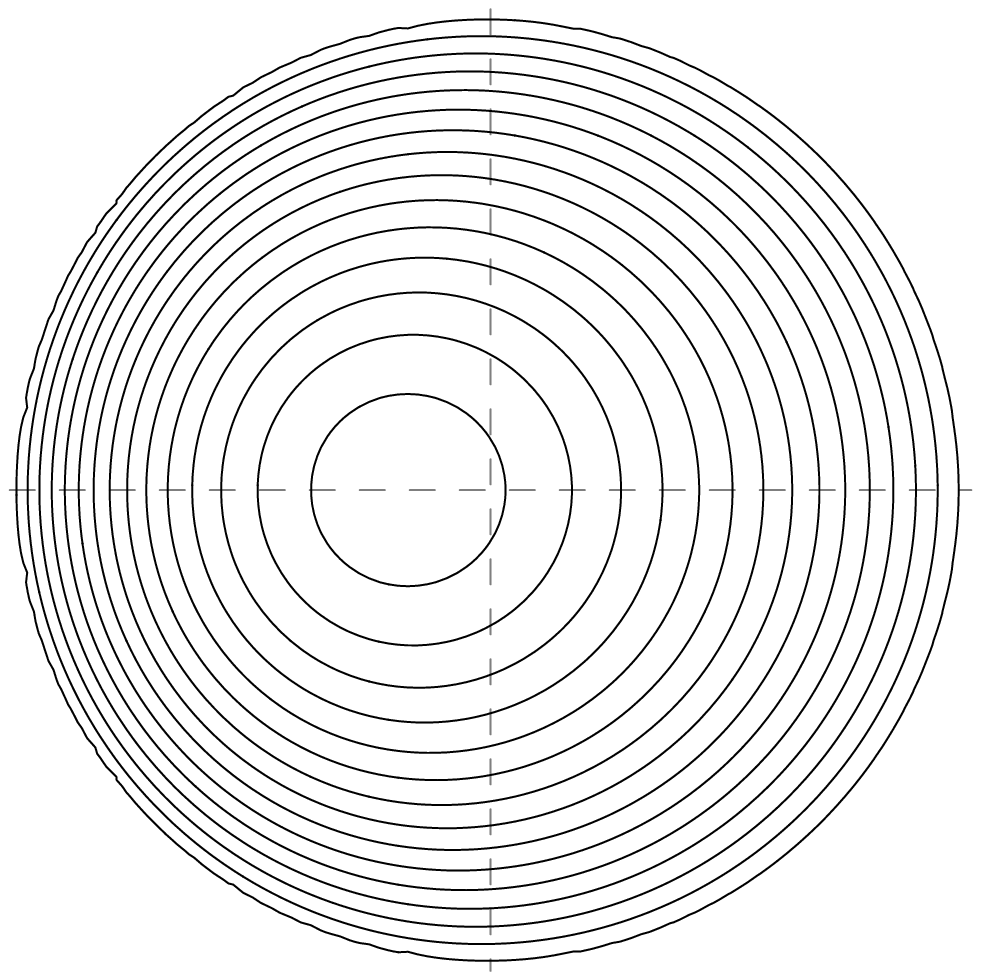}}
\put(110,100){\includegraphics[width=0.2\textwidth]{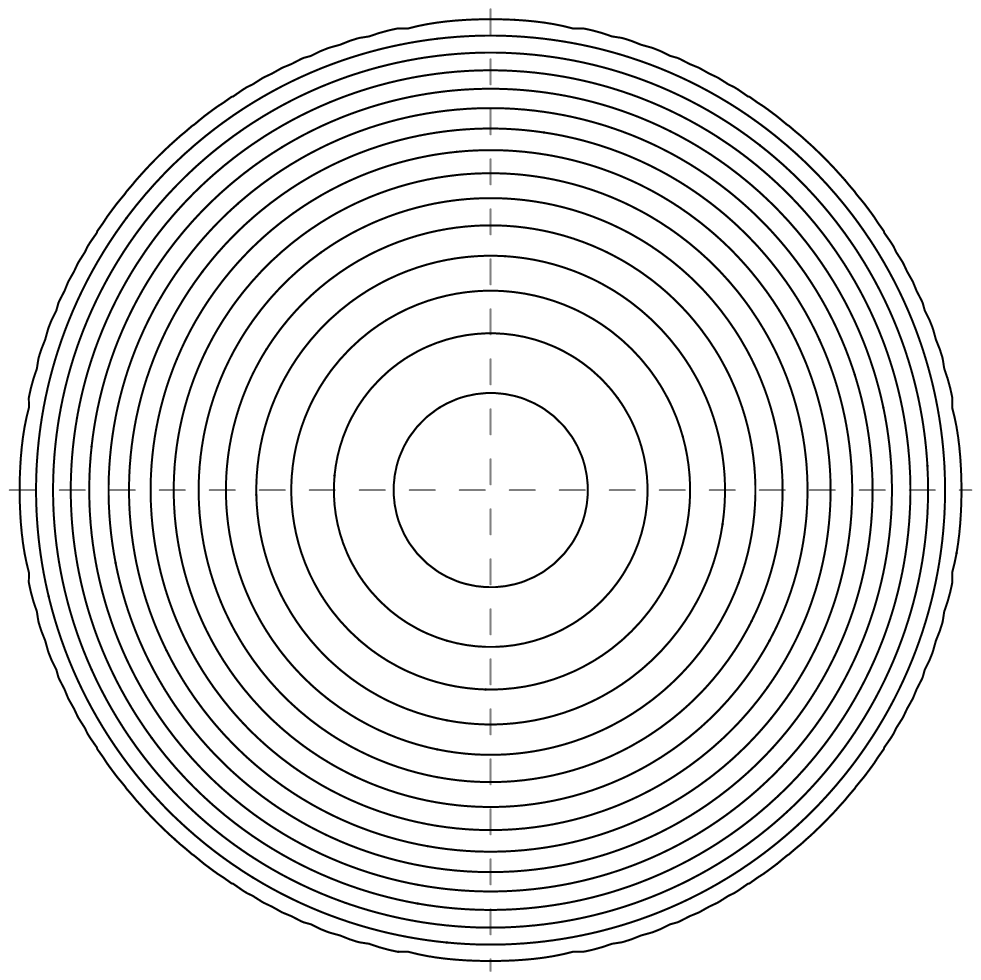}}
\put(0,0){\includegraphics[width=0.2\textwidth]{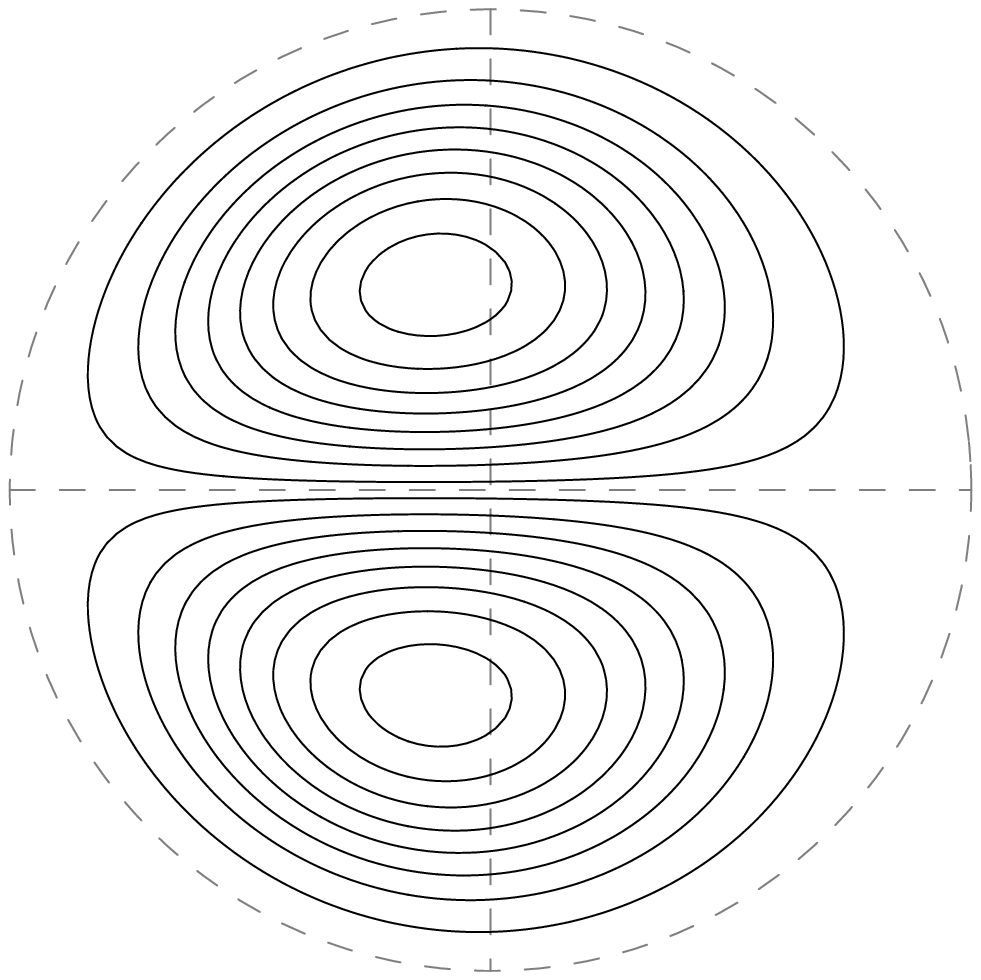}}
\put(110,0){\includegraphics[width=0.2\textwidth]{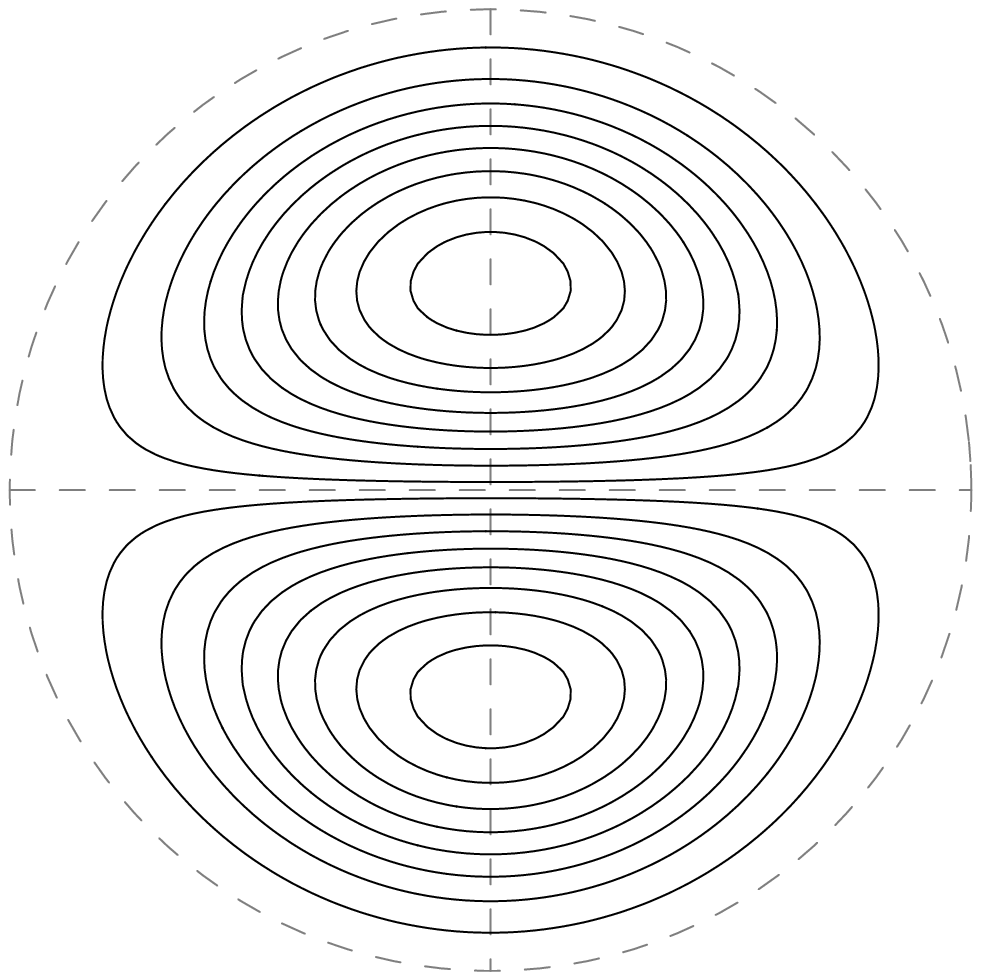}}
\put(7,193){(a)} \put(117,193){(b)} \put(7,93){(c)}
\put(117,93){(d)}
\end{picture}
\caption{$\kappa=0.5$, $\R=1$, $F=0$. Comparison of solutions. (a)
$v_\zeta$, (b) $v_\zeta^{(c)}$, (c) $\sfunc$, (d) $\sfunc^{(c)}$.}
\label{fig_c1}
\end{figure}

\begin{figure}
\includegraphics[width=0.35\textwidth]{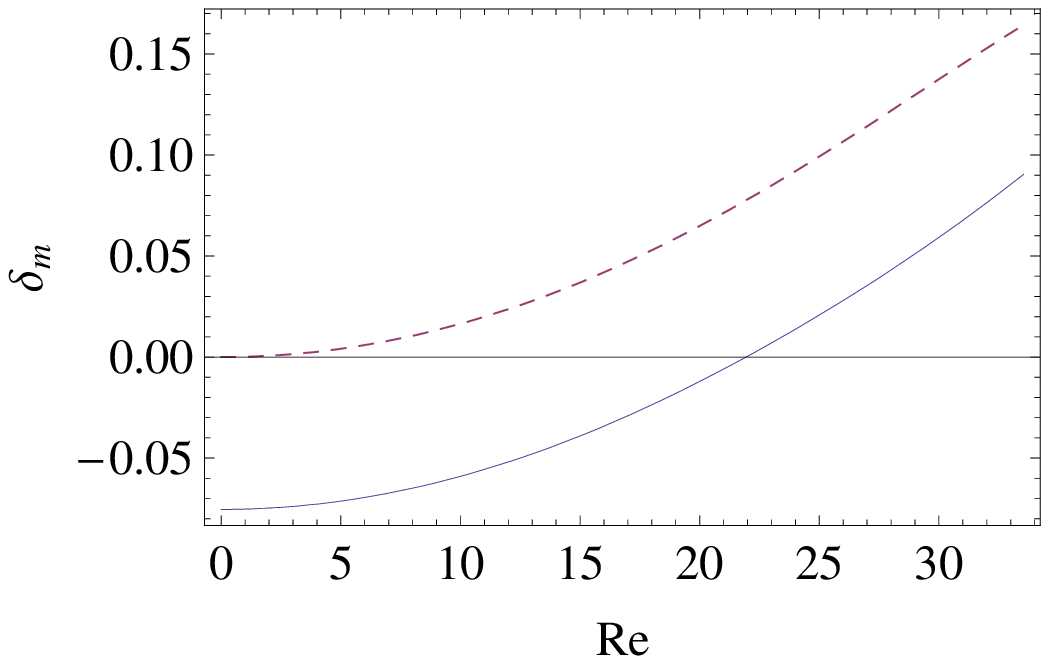}
\caption{Shift $\delta_m$ of the maximum of stream velocity vs $\R$
at $\kappa=0.1$. Dashed line stands corresponds to the solution
$v_\zeta^{(c)}$.}\label{shift_pic}
\end{figure}

In the case of $F=0$ the result is shown in Fig.~\ref{fig_c1}.
One can see that the shift $\delta_m$ of the stream velocity maximum
due to a centrifugal force directed outward from the torus axis is
not valid at small Reynolds number. In fact for $\kappa\gtrsim 0.3$
there is a shift {\it only} in  the opposite direction (inwards)
because of viscous stresses in the region of convergence. This shift
is mainly due to the difference \eqref{new_v_term}. It has a
dependence on $\R$ shown in Fig.~\ref{shift_pic}.

\begin{figure}
\begin{picture}(200,100)(0,0)
\put(0,000){\includegraphics[width=0.2\textwidth]{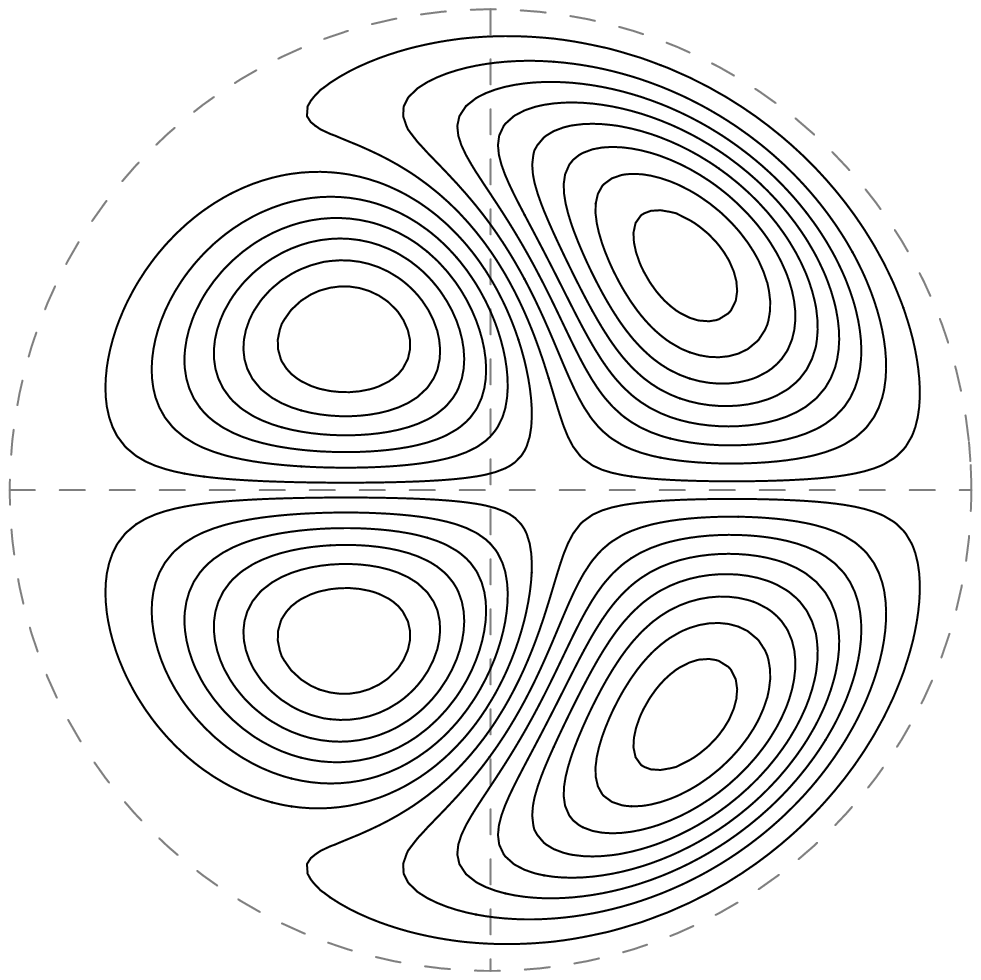}}
\put(110,000){\includegraphics[width=0.2\textwidth]{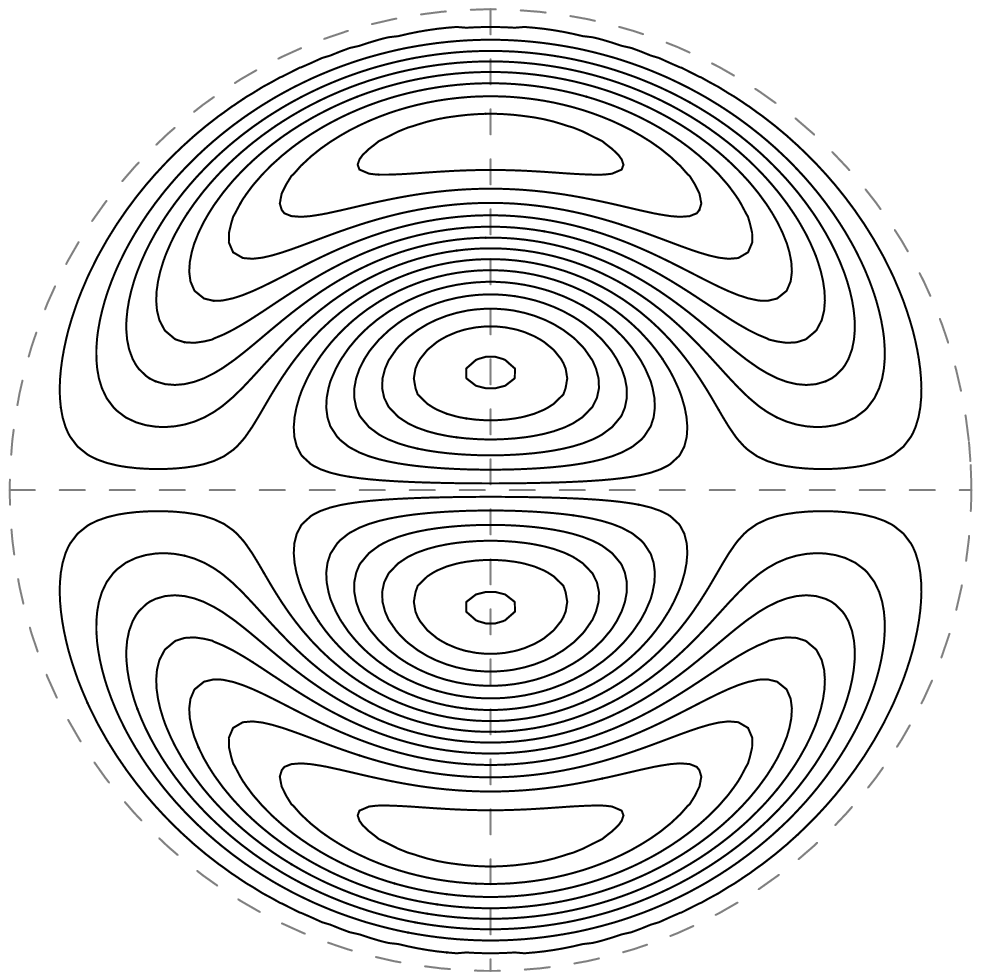}}
\put(7,93){(a)} \put(117,93){(b)}
\end{picture}
\caption{$\kappa=0.5$, $\R=1$, $F=-0.63$. Comparison of solutions.
(a) $\sfunc$, (b) $\sfunc^{(c)}$.} \label{fig_c2}
\end{figure}

When the torus rotates ($F\neq0$) the Coriolis force gives rise to
additional vortices in the cross-section. In the counter-rotating
case such a vortex can act against the centrifugal vortex. This
produces a four-vortex picture (see Fig.~\ref{fig_c2}). Again we see
that the flow pattern is different for a low Reynolds number. The
vortex corresponding to the Coriolis force arises at the boundary
while $\sfunc^{(c)}$ starts to grow in the center. \nocite{Dyke1975}

\begin{figure}
\includegraphics[width=0.40\textwidth]{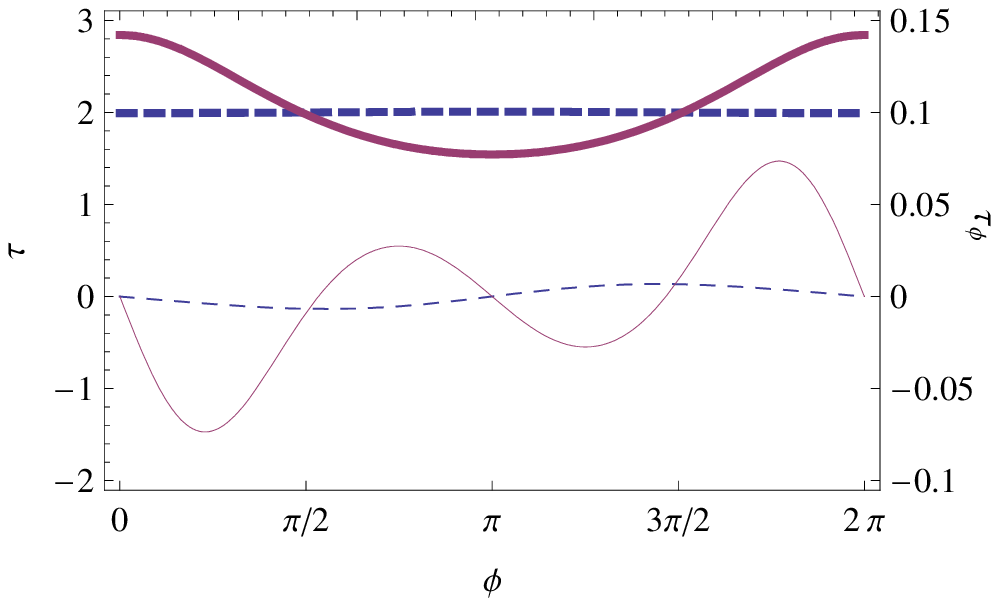}
\caption{Axial shear-stress $\tau$ (bold lines, tick labels at left)
and azimuthal shear-stress $\tau_\phi$ (thin lines, tick labels at
right) dependencies on angle: Dashed lines corresponds to the
solution in \cite{zhang:056303}. $\kappa=0.4, {\R}=10, F=-0.49$.}
\label{stress}
\end{figure}

Axial shear-stress along the channel
$\tau=-\left.\pd{v_\zeta}{\rho}\right|_{\rho=1}$ differs from that
presented in \cite{zhang:056303}. The first order residual is $3/16
G \kappa\left(3 \rho ^2-1\right) \cos\phi$. It does not change the
pressure drop but produces a strong variation (about 40\%) of
friction at the boundary (see Fig.~\ref{stress}, thick curves).
There is a similar difference in the azimuthal shear-stress
$\tau_\phi=-\left.\pd{v_\phi}{\rho}\right|_{\rho=1}$. The second
order residual is
\begin{equation}\label{az_shstres_dif}
\tau_\phi-\tau_\phi^{c}=\frac{G  \R(13 G-25) \kappa ^2 \sin 2
\phi}{3840}.
\end{equation}

This work shows that solution of the {\it full} governing equations
reveals some specific features for flow in a toroidal channel: a
shift of the maximum of the stream velocity toward the inner axis,
the appearance of a second pair of vortices at the internal
boundary, an additional $\phi$-dependance of stresses. These are
well pronounced at low values of the Reynolds number. The solution
for a giving curvature $\kappa$ asymptotically approaches the known
solution \cite{zhang:056303} at high $\R$.

\begin{acknowledgments}
This work is supported by RFBR-Ural grant No.~06-01-00234 and the
Russian Federation President grant MK-4338.2007.1.
\end{acknowledgments}

\end{document}